# Experimental electronic stopping cross section of tungsten for light ions in a large energy interval

M. V. Moro[1, a], P. M. Wolf[1], B. Bruckner[1], F. Munnik[2], R. Heller[2], P. Bauer[3] and D. Primetzhofer[1]

[1]Department of Physics and Astronomy, Uppsala University, Box 516, 751 20 Uppsala Sweden
[2]Helmholtz-Zentrum Dresden-Rossendorf, Institute of Ion Beam Physics and Materials Research, Bautzner Landstraße 400, 01328 Dresden Germany
[3]Atomic Physics and Surface Science, Johannes Kepler University, A-4040 Linz Austria

**Abstract**

Electronic stopping cross section of tungsten for light ions was experimentally measured in a wide energy interval (20 to 6000 keV for protons and 50 to 9000 keV for helium) in backscattering and transmission geometries. The measurements were carried out in three laboratories (Austria, Germany and Sweden) using five different set-ups, the stopping data deduced from different data sets showed excellent agreement amongst each other, with total uncertainty varying within 1.5 - 3.8 % for protons and 2.2 - 5.5 % for helium, averaged over the respective energy range of each data set. The final data is compared to available data and to widely adopted semi-empirical and theoretical approaches, and found to be in good agreement with most adopted models at energies around and above the stopping maximum. Most importantly, our results extend the energy regime towards lower energies, and are thus of high technological relevance, e.g., in fusion research. At these low energies, our findings also revealed that tungsten - featured with fully and partially occupied f- and d-subshells, respectively – can be modeled as an electron gas for the energy loss process.

*Keywords:*
*Stopping power; Tungsten; Free electron gas; Bragg peak; Protons; Helium; Fusion.*

[a] Corresponding author.
E-mail address:  marcos.moro@physics.uu.se

1. Introduction

When an energetic ion penetrates matter, a retarding force acts on the ion as it starts losing energy due to collisions with target electrons and nuclei. The average energy loss of the ion per unit path length is called stopping power, *S*. This quantity has been under scientific investigation for over a century [1] and is nowadays of utmost importance in practically any technological or scientific application involving charged particles, like ion beam material modification and analysis [2,3], semiconductor industry [4,5], hadron-therapy [6,7] and fusion research [8,9]. Ion-solid interaction has also provided notable insights into solid state physics [10,11] and, using sufficiently slow ions, into non-equilibrium physics [12–14].

For low energy ions, i.e., when the ion velocity *v* is comparable or smaller than the Fermi velocity $v_F$ of the target ($v \leq v_F$), the electronic energy loss is dominated by excitation of valence electrons. Thus, a simple but powerful model of electronic energy loss describes the target electrons as Free Electron Gas (FEG) [15]. This approach predicts *S* to be proportional to the ion velocity, so that $S = Q(Z_1, r_s) \cdot v$. The friction coefficient *Q* may be calculated using, e.g., density function theory [16] and depends on the projectile atomic number $Z_1$ and on the one-electron Wigner-Seitz radius $r_s = (3/4\pi n_e)^{1/3}$ ($n_e$ is the FEG density corresponding to the number of valence electrons $N_{val}$ per atom of the material) [17]. At high ion energies (i.e., $v \gg v_F$), the ion can be considered only a weak perturbation of the target electronic system, and the process is commonly modeled by first-order theories [18–20]. To avoid a trivial dependence on the material mass density, it is convenient to normalize *S* by the atomic density *n* of the material. The resulting quantity is commonly referred to as stopping cross section, $\varepsilon = S/n$.

We report experimental electronic stopping cross sections of W for light ions in a wide energy range (20 keV to 6 MeV for protons and 40 keV to 9 MeV for helium). We have deduced data from energy spectra recorded in both backscattering (BS) and transmission (TR) geometries. The choice of W as the material of interest is motivated by several facts. W is expected to form a major plasma facing component (PFC) in the divertor of next generation fusion devices. To model the expected material modification knowledge on energy deposition by plasma species is urgently required, but no data at low energies is available. Also, ion beam analysis as a commonly employed tool in characterization of PFC's [21] requires accurate reference data. Finally, W features a complex electronic structure with a shallow but complete f-subshell and half-filled d-states which yields information complementary to recent studies indicating the inapplicability of a FEG-model to describe electronic stopping in early transition and rare-earth metals [22]. We also compare our obtained results to semi-empirical (ICRU [23] and SRIM [24]) and *ab-initio* approaches (Montanari et al. [25], CasP [26] and DPASS [27]).

**2. Experimental approach**

2.1 Sample preparation

Two different types of tungsten samples, i.e., bulk and self-supporting foil targets were selected for the present work. Backscattering (BS) experiments were performed on a commercially available (MaTecK GmbH, 99.9 % nominal purity) polycrystalline tungsten foil (0.5 mm thick and cut in 10x10 mm$^2$). Prior the analysis, the sample was cleaned with acetone followed by isopropanol (99 %) in ultrasonic bath. As information on the possible presence of impurities on and in the sample is essential for the reliability of the final deduced stopping data [28], the sample quality was repeatedly assessed by coincidence time-of-flight-energy heavy elastic recoil detection analysis (ToF-E ERDA), using 36 MeV I$^{8+}$ as probing beam at Uppsala University (information on the technique and data analysis are found elsewhere [29]). ToF-E ERDA depth-profiles (not shown) indicated no significant contaminations in the bulk of the sample (quantification limit around 0.1 at.% for the present technique), apart from superficial contaminations of H, C and O (≈ 5 at.% in total, within 20 nm), most likely due to adsorbed species from exposure to air.

For the transmission experiments, we manufactured ≈ 50 nm thick self-supporting tungsten foil (i.e., without substrate). The fabrication process consisted of thermal evaporation of a NaI salt as buffer layer onto Si wafer, followed by a W thin film magnetron sputtered on top. The inorganic salt is thereafter dissolved in distilled water, and the floating W foil is carefully captured on an annular sample-holder suitable for the ToF-MEIS set-up (discussed below). The areal density of the tungsten is quantified by 2 MeV He$^+$ Rutherford Backscattering Spectrometry (adopting stopping powers deduced from our backscattering measurements). The contaminants for sample used in this work (15 at.% O, 9 at.% C and 5 at.% H) are determined via ToF-E ERDA analysis, as described above, and, for some foils, by Elastic Backscattering Spectrometry, employing the $^{16}$O($\alpha,\alpha_0$)$^{16}$O resonant reaction at 3.037 keV helium [30,31] – this latter technique is used for a better depth-resolved oxygen quantification. A more detailed description of the manufacturing process and sample quality is given elsewhere [32].

2.2 Experimental set-ups

Experimental backscattering and transmission measurements were carried out using in total five set-ups. Thus, we cover a broad energy range and are able to check the consistency of the final results in overlapping energy ranges, decreasing thus possible systematic uncertainties arising from e.g., the adopted experimental technique and energy calibration of the accelerator. For this reason, not only different set-ups and energy ranges were selected but also different geometries were adopted. Low energy backscattering measurements were carried out at the AN-700 van de Graaff accelerator of Johannes Kepler University (JKU) [33] in Austria for D[+] and H[+] ions, and at the new multi-purpose set-up for low-energy ion scattering assembled at the third beam line (BL3) of the 350-KV Danfysik implanter at Uppsala University (UU) [34] in Sweden, for H[+] and He[+] ions. Additionally, measurements in both geometries, BS and TR, were performed using the ToF-MEIS system [35] at the implanter (UU) for helium as projectile. Additional measurements in BS geometry for protons and helium at energies around 1 MeV were carried out at the µ-beam line of 3 MV Tandem Accelerator at the Helmholtz-Zentrum Dresden Rossendorf (HZDR) in Germany. Finally, high energy stopping data were deduced from backscattering measurements conducted at the 5 MV NEC 15SDH 2 Tandem accelerator of Uppsala University using H[+] and He[+,2+] as probing species [21]. A summary of employed laboratories, set-ups (main scattering chamber), samples (bulk or self-supporting foil), methods (backscattering or transmission), projectile primary energy ranges, geometries and detector resolutions is given in Table 1.

**Table 1.** Summary of experimental conditions (i.e., set-ups, samples, experimental method, energy range, geometries, and detector resolution) used in this present work to deduce the electronic stooping data of W.

| Laboratory – city (Set-up / Accelerator) | Sample / Method[*] | Primary ion energy range [keV] | | | Geometry[**] [°] | | Detector[***] FWHM [keV] | |
|---|---|---|---|---|---|---|---|---|
| | | D[+] | H[+] | He[+] | α | θ | H[+] | He[+] |
| JKU – Linz (RBS / AN-700 Accel.) | W-bulk / BS | 80-160 | 100-400 | --- | 0 | 154.1 | 8 | 9 |
| UU – Uppsala (BL3 / Implanter) | W-bulk / BS | --- | 20-250 | 50-300 | 15 | 135 | 4 | 7 |
| UU – Uppsala (ToF-MEIS / Implanter) | W-foil / TR | --- | --- | 80-200 | 0 | 0 ± 1 | --- | 1-2.5 |
| UU – Uppsala (ToF-MEIS / Implanter) | W-foil / BS | --- | --- | 80-200 | 0 | 120 | --- | 1 |
| UU – Uppsala (ToF-MEIS / Implanter) | W-bulk / BS | --- | --- | 40-100 | 5 | 160 | --- | 1.5 |
| HZDR – Dresden (µBeam / 3 MV Tandem) | W-bulk / BS | --- | 500-2200 | 500-2200 | 0 | 172 | 16 | 18 |
| UU – Uppsala (T4-1 / 5 MV Tandem) | W-bulk / BS | --- | 2000-6000 | 2000-9000 | 5 | 170 | 13 | 14 |

[*] BS and TR are, respectively, backscattering and transmission geometry (see text and references therein).
[**] α and θ are incident and scattering angles with respect to sample normal and beam direction, respectively.
[***] Detector resolution at FWHM in keV and averaged over the respective energy chain.

2.3 Evaluation procedures

In the backscattering approach, apart from the experiments employing self-supporting foils, we rely on the fact that the height $H$ of an energy spectrum contains information on the stopping cross section factor $[\varepsilon]$, which in turn, includes the electronic stopping cross section $\varepsilon$ of the projectile on the way in and on the way out of the target [36,37]. Direct evaluation of the electronic stopping from a single BS energy spectrum would require precise and accurate knowledge on the relevant experimental parameters [38,39]. Therefore, we performed relative measurements between the sample of interest, i.e., W to a reference sample with well-known stopping power, i.e., Au under otherwise identical experimental conditions [40]. Thus, the ratio of the heights of the experimental spectra, $H_{W,expt}/H_{Au,expt}$, contains information on the stopping cross section ratio $\varepsilon_{Au,expt}/\varepsilon_{W,expt}$. For a quantitative SCS evaluation, $H_{W,expt}/H_{Au,expt}$ is compared to the ratio obtained from simulations, $H_{W,sim}/H_{Au,sim}$, using either the Monte Carlo TRBS [41] or the SIMNRA [42] codes. Such simulations account for potential differences in cross sections due to screening, multiple scattering (of higher interest towards lower energies). Agreement between height ratios of experimental and simulated spectra is obtained by variation of the W stopping power used as input in the simulation, as the only adjustable parameter [43]. Note that our BS approach also ensures acquisition of both spectra for the same primary ion charge. The energy interval of evaluation is chosen close to the kinematic onset of the spectra, reducing thus multiple scattering effects and ensuring sufficient linearity of the specific energy loss. We selected Au as reference material due to its vicinity in atomic number (hence kinematic factor), easy manufacturing process (guaranteeing high purity bulk samples), and due to its abundancy in well-established data sets [44–47] selected from the IAEA database [48]. A detailed description of the present approach (considering similar energy regimes), is given by the authors elsewhere [49]. Fig. 1 (a) shows typical relative backscattering spectra of Au (black dotted line) and tungsten (red dotted line), recorded for 190 keV He[+] in at BL3 set-up of implanter (UU) (see Tab. 1). The agreement between experiment and simulation with optimized $\varepsilon$ (solid lines) is excellent within the evaluated energy window (vertical blue dashed lines). To demonstrate the sensitivity of the spectrum height to the stopping power, simulations with $\varepsilon_W$ modified by ± 5 % are also presented (see red dashed lines in Fig. 1 (a)).

For the transmission experiments, the sample-holder with a self-supporting W foil (see Sec. 2.1) is mounted in the UHV ToF-MEIS set-up at the implanter (UU) in normal incidence with respect to the beam direction (i.e., $\alpha = 0°$). A large solid angle (130 msr) MCP/delay-line position sensitive detector [50], which can be circularly rotated around the sample, is placed with its center at 0° (and evaluation radius of ±1°) in transmission position with respect to the beam direction. The energy loss $\Delta E$ is calculated from the TOF shift between primary beam and particles transmitted through the foil. The SCS is evaluated using the mean energy approximation [37], which is proved to be accurate enough for the present energy range [51] (texture effects in the foil are negligible). Monte Carlo simulations using TRBS are performed to account for effects of multiple scattering, and as in BS $\varepsilon_W$ is adjusted so that the simulations reproduce the experimental data. In these simulations we have also considered the foil impurities, as they may influence the multiple scattering calculations (especially towards lower energies [52]). The final W stopping data is extracted from the simulation by employing Bragg's rule [53]. In Fig. 1 (b), we show a typical experimental spectrum of 100 keV helium ions transmitted through a tungsten foil of (50 ± 1) nm (black dotted line), as well as a simulation with optimized stopping power $\varepsilon_W$ (red solid line). The small peak at $E_0$ in panel (b) featured particles that directly hit the detector, due to pinholes in the foil target. By simultaneous detection of both peaks, possible systematic fluctuations in either beam energy and/or detection electronics - even if small - are canceled out in the evaluation. The result of the simulation is convoluted with a Gaussian to account for the finite energy resolution of the detector and straggling.

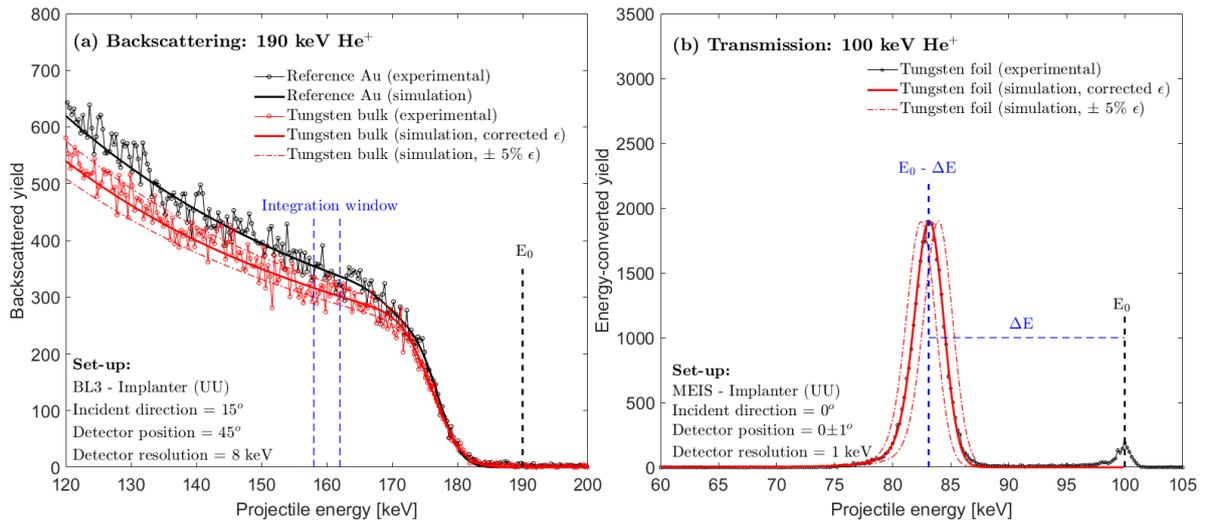

Figure 1 (Color online). (a) Experimental spectra for 190 keV He$^+$ backscattered from W (black dotted line) and Au (red dotted line) recorded at the BL3 set-up (UU – see Tab. 1) for an incident and detection angle of $\alpha$ = 15° and θ = 135°, respectively. Simulations using SIMNRA are also depicted (solid lines for W and Au). (b) Experimental transmission spectrum for the self-supporting W foil recorded at the ToF-MEIS setup (UU – see Tab. 1), deploying 100 keV He$^+$ at $\alpha$ = 0° and θ = 0 ± 1° (see text for details). The red solid line is the result of a Monte Carlo simulation using TRBS. To demonstrate the sensitivity of both evaluation procedures, the corrected $\varepsilon_W$ is varied within ± 5 % (red dashed lines).

## 3. Results and discussions

In Fig. 2 (a), the experimentally deduced stopping cross section of W (blue filled symbols) for protons in the energy range between 20 keV and 6 MeV is shown: left and right triangles are results obtained with D$^+$ and H$^+$ using the 700 keV accelerator at JKU (Austria), respectively, circles are data from the implanter using BL3 at UU (Sweden), diamonds are data from the 3 MV Tandem at HZDR (Germany) and, finally, squares are results from 5 MV Tandem at UU (Sweden). See Tab. 1 for more details. Our results for protons feature a total uncertainty (i.e., random and potential systematic contributions) of ≈ 1.5 - 3.8 %. For example, the low energy (protons) data set corresponding to measurements at BL3 (see Tab. 1) show an averaged uncertainty of ≈ 3.8 %, whereas high energy stopping data at T4-1 were deduced with an (averaged) uncertainty of ≈ 1.5 %. The final uncertainty was calculated following recommendations given in [54,55], and taking into account uncertainties from: i) integral beam dose (i.e., for Au and W spectra), ii) pile-up contributions and iii) reference data (for BS method), as well as iv) counting statistics (BS: spectra ratio and TR: peak position) and v) fit precision (BS and TR). For a more detailed description of how each contribution affects the deduced stopping data, we refer to Sec. 5 of [49]. To present the stopping cross section in a continuous and smooth manner, and to avoid problems with data interpolation, we fitted our $\varepsilon_W$ values using the Ziegler-Biersack (ZB) function, as described in [56] (see blue solid line). The averaged fit precision was found to be ≈ 1.5 %. The agreement between different data sets is evident (blue filled symbols), ensuring consistency in the experimental and data evaluation approaches. Our results are compared also to data from literature [48] (black open symbols). In total, 65 data points from 4 different data sets [57–60] are shown in panel (a), with stated uncertainties varying within 3-10 % (black open symbols). In general, literature data agrees well with our results, within the uncertainties and energy range ≈ [80-6000] keV.

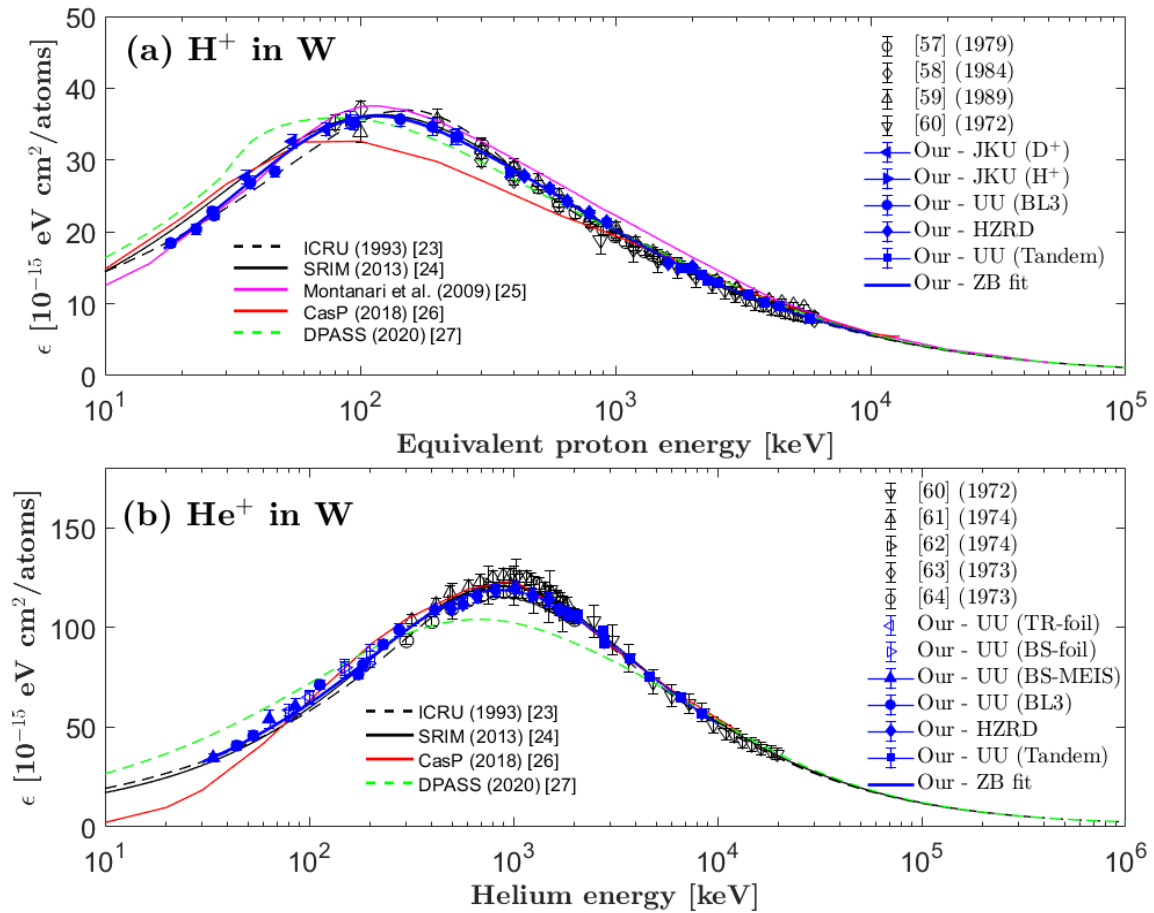

Figure 2 (Color online). (a) Experimental stopping cross section of W for protons (filled symbols) and available data found in literature (open symbols). Comparison to semi-empirical and theoretical models is also depicted (continuous and dashed lines – see legend). (b) Same as in panel (a), but for helium. See text for details about data sets and models.

In Fig. 2 (b), the experimental $\varepsilon_W$ data is shown for He in the energy range 40 keV to 9 MeV (blue filled symbols): left and right triangles are data deduced from TR and BS measurements in the ToF-MEIS set-up at UU, respectively, using the self-supporting W foil, while upward triangles are results derived from BS measurements in the ToF-MEIS set-up, but using the bulk W sample. Circles, diamonds and squares are stopping values obtained from BS measurements using BL3 at UU, micro-beam line at HZDR and T4-1 at UU, respectively. Similarly, as for protons, the total uncertainties are estimated to be ≈ 2.5 - 5.5 %, (see comments for protons). The ZB fit to our data, shown as blue solid line, has a precision of ≈ 1.7 % over the entire energy chain. Again, excellent agreement amongst different set-ups, methods, samples and geometries is observed. Experimental helium data is compared to literature available in [48] (black open symbols): 62 data points from 5 different data sets [60–64] are shown in Fig.2 (b), with stated uncertainties of 2 to 8 %. In general, good agreement between our data and literature values is achieved. Around the maximum stopping (i.e., ≈ 1 MeV), there is a systematic difference of up ≈ 6 % between the data from [61] (open upward triangles) and [64] (open circles); our data better confirm the former within ≈ 1 %.

We also compared the experimental $\varepsilon_W$ data of this work and from literature with several semi-empirical and *ab-initio* models (other lines in Fig. 2). In the former case, we employed two widely used semi-empirical approaches: the last released version of SRIM (2013) [24] and the online compiled tables for protons [65] and helium [66], which are based on the International Commission on Radiation Units and Measurements (ICRU) Report 49 [23]. Both approaches are similar in their underlying physics: the stopping power at higher energies (typically > 1 MeV/u) is generally evaluated based on a Bethe-Bloch-like formalism, whereas at lower energies (< 1 MeV/u), fitting formulas are adopted, while interpolations between these two energy regimes is usually done adopting Fano slope-plots [67,68]. Therefore, the precision of such evaluations effectively depends on the availability and reliability of the experimental data. One important advantage of such semi-empirical approaches is that their outputs can always be improved in new releases by updating their internal databases.

For the *ab-initio* stopping predictions, we used three different approaches. First, in the formalism by Montanari et.al. (only for protons), the electronic W output (retrieved from [25]) is calculated from two main contributions: i) from the bound electrons, responsible for the energy loss in the high energy regime using the Shell Wise Local Plasma Approximation (SLPA) [69], and ii) from the valence electrons, responsible for the energy loss at lower energies (in the FEG regime). In the second approach, the CasP code [26] is used to predict stopping values by employing the Unitary Convolution Approximation (UCA) [70] for energies in the Bethe regime and Transport Cross Section (TCS) [71,72] model for the low energy regime. Finally, the third approach consisted of adopting the recent version of DPASS tabulations (2020) [27], which evaluates the stopping power in the Binary theory [73], which is based on Niels Bohr's classical approach [74], but with improved modelling of the underlying physics (see Ref. [75] and discussions therein).

To assess in more details the capability of the each model to reproduce our experimental data in the investigated wide energy range, we show in Fig. 3 the difference between the respective models to our experimental data ( $\frac{\varepsilon^{Model} - \varepsilon^{ZB-fit}}{\varepsilon^{ZB-fit}} \times 100\%$ ) for protons (a) and helium (b).

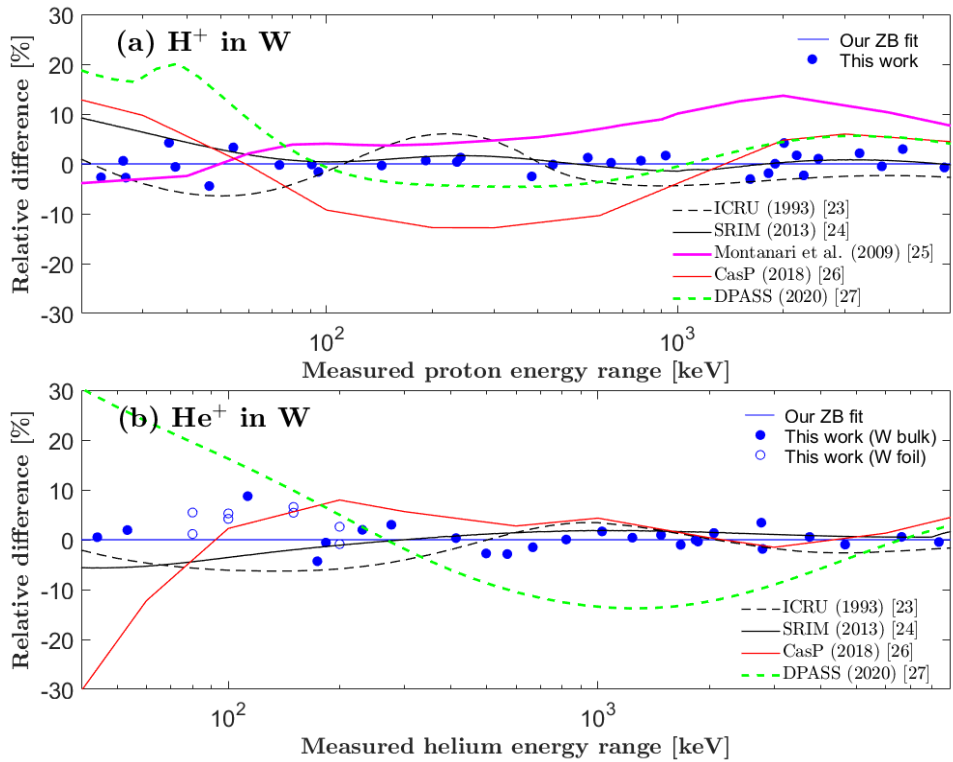

Figure 3 (Color online). Difference (in %) between the model (see color dashed and solid lines in the legends) and our data (represented by the blue horizontal continuous line) within the respective measured projectile energy range: [20 - 6000] keV for protons (a) and [40 - 9000] keV for helium (b).

In this figure, the blue horizontal solid line is equivalent to a prediction in agreement with our ZB fit, as discussed above (other color lines represent the models - see respective legends). As one can see in Fig. 3, the semi-empirical approaches are most capable of reproducing our data from intermediate (here: 100 keV ≤ E ≤ 1 MeV) to high energies (here: E ≥ 1 MeV) regardless of the projectile. At intermediate energies, ICRU (black dashed line) and SRIM (black solid line) agree to our data within 3.5 % and 1.5 % for protons, respectively and within 4 % and 2 % for helium, respectively. At higher energies, our data show better agreement with SRIM (∼ 1 % for proton and helium) than with ICRU (2 % and 2.5 % for proton and helium, respectively). Nevertheless, ICRU stopping seems to oscillate as a function of energy from intermediate to low regions (here: E ≤ 100 keV) within ± 5 %. These oscillations may be related to interpolation issues and extrapolation to regions without experimental data available. In the low energy regime (without available data), both ICRU and SRIM feature larger deviations from our data of up to ∼ 6 % both for protons and helium, on average.

The prediction of the energy loss for protons according to Montanari et al. (magenta solid line) agrees within ~ 4 % to our data for energies around the stopping maximum stopping (~ 150 keV), whereas towards higher energies, predicted values exceeds measured data by ~ 10 % The fact that SLPA predictions for the stopping power are high with respect to literature data and to predictions by the Bethe-formalism, has already been identified by the authors also for other elements from period 6 (Pb and Bi) [76]. On the other hand, their formalism holds a good precision in the low energy regime (FEG zone), within ~ 2-3 %. The CasP model (red solid line) yields better precision for helium than for protons at high (3 % and 5 %) as well as intermediate (4 % and 10 %) energies, respectively. However, towards low energies, the discrepancy between CasP and experiments increases up to ~ 9 % for protons and up to ca. - 30 % for helium. The DPASS code (green dashed line) overestimates our data for protons within ~ 20 % at the lowest energy, then it improves back to ~ 5 % in the intermediate and high energy regimes. For helium, the situation is similar: starting at ~ + 30 % in the FEG regime predictions become lower than experimental data down to - 20 % in the intermediate energy zone, and then increases again up to 2 % at the highest energies. Altogether, the decreasing accuracy in predictions by SRIM as a commonly employed source of tabulated extrapolations represents a challenge for modelling of low-energy ion irradiation effects [77,78], which can be at least partially overcome by additional experiments at significantly lower energies.

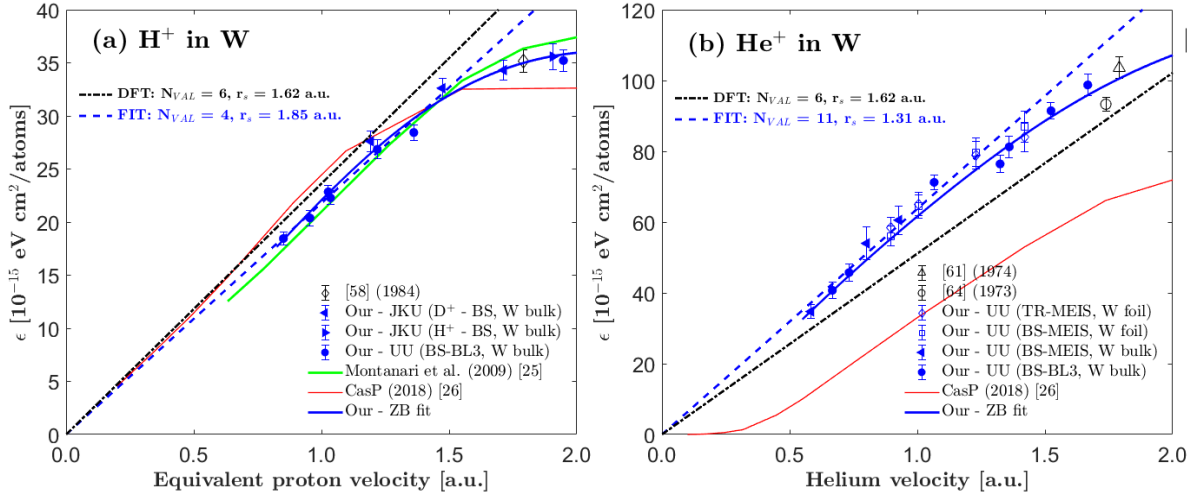

Figure 4 (Color online). Experimental stopping cross section of W (blue solid and empty symbols) plotted as a function of proton (a) and helium (b) velocities. The blue solid line represents our ZB-fit (discussed previously), the green solid line is the Montanari et al FEG approach (panel (a)), whereas the red solid line represents stopping prediction by the CasP. The blue dashed line is the linear fit to our data. The black dash-dotted line is DFT calculation [79] assuming $N_{val}$ ≈ 6 electrons (see text for details).

In Fig. 4, the electronic stopping cross section of W is depicted in a different format, i.e. as a function of ion velocity (up to 2 a.u.) for protons (panel (a)) and helium (panel (b)). Blue open and solid symbols are data from this work, while the only one stopping data point available in literature for protons in the velocity range shown in this figure is presented as black open diamond (panel (a)). As one can see, at ion velocities ≤ 1 a.u. data is proportional to the ion velocity. The experimental friction coefficient $\varepsilon/v$ is obtained for both projectiles by fitting the data (blue dashed line) for further comparison with theory.

Using density function theory (DFT) calculations from [79,80] for the proton data (panel (a)), we obtain $r_s$ ≈ 1.85 a.u., corresponding to a valence electrons density $N_{val}$ ≈ 4,. For comparison, assuming the electronic configuration of tungsten as [Xe]$6s^2 4f^{14} 5d^4$, we calculated the friction coefficient assuming $N_{val}$ = 6 electrons corresponding to $r_s$ ≈ 1.62 a.u. (black dash-dotted line), which overestimates the experimental stopping data by ∼ 10 %. We found identical behavior for the CasP code (red solid line) as expected, since CasP assumes the same FEG electron density (its default option is $r_s$ ≈ 1.62). Note that the experimentally deduced plasmon energy for tungsten has been determined to be 24.3 eV [81] corresponding to $r_s$ ≈ 1.55 a.u ($N_{val}$ ≈ 6.8). On the other hand, the formalism as described by Montanari et al., reproduces our data within good precision (∼ 1-2 %). Even though these authors have also considered $N_{val}$ ≈ 6 electrons [25], they assumed some energy vicinity effects between the 5p and 4f subshells, leading to an inter-shell screening effect that needs to be considered in their FEG model (see [25] and discussions therein). Concordantly, a possible influence of f-electrons on electronic stopping of H ions has also been discussed for proton stopping experiments in other transition (Pt) and rare-earth (Gd) metals in the low energy regime [22], and more recently, even at energies up to the stopping maximum (i.e., Bragg peak) for several early transition metals [82]. For helium ions (panel (b)), the situation is significantly different. Even though velocity proportionally is observed (blue dashed line), for the DFT calculations [79] to match the experimental data $r_s$ ≈ 1.31 a.u. would be required, which yields $N_{val}$ ≈ 11 electrons. On the other hand, if we compare a DFT model assuming $N_{val}$ ≈ 6 electrons, the results underestimate our data by ∼ 20 % (see black dash-dotted line). This observation can be explained by the fact that for dressed ions Coulomb excitation of electron hole pairs in an otherwise unperturbed electronic system is not the only excitation mechanism, and other additional energy loss channels are open. Such a behavior has also been observed for electronically simple systems such as He in Al [83] and explained by charge exchange [84], and effects of projectile excitation and charge state on the interaction potential [85,86]

## 4. Summary and conclusions

In this work, we have experimentally evaluated the electronic stopping power of W in a wide energy range for protons (from 20 keV to 6 MeV) and helium (from 40 keV to 9 MeV) from energy spectra recorded in backscattering and transmission geometries. Measurements were carried out in five different set-ups from three laboratories (Austria, Germany and Sweden) in various energy ranges, geometries and using different samples (i.e., foils and bulks). The agreement amongst the data sets is excellent, and the final stopping data is obtained with a total averaged uncertainty (random and systematic) between ∼ 1.5 - 3.8 % for protons and ∼ 2.5 - 5.5 % for helium. Our results represent an improvement in the availability of experimental data, especially towards low energies (i.e., FEG regime), which are of particular interest for e.g., fusion research, where no data is available.

Our stopping results have also been systematically compared to the mostly used semi-empirical (ICRU and SRIM) and *ab-initio* theoretical approaches (Montanari et al. formalism for protons and CasP and DPASS for protons and helium). We demonstrated that the precision of the model strongly depends on the energy range of interest: for high (≥ 1 MeV) and intermediate (100 keV ≤ E ≤ 1 MeV) energies, the ICRU and SRIM results demonstrate best agreement to our data, with precision better than ∼ 4 % in average for both projectiles. Towards the low energy regime (E ≤ 100 keV, FEG regime), the approach by Montanari et al. are closest to our experimental data (∼ 2-3 %.). In agreement with earlier observations for transition metals, our findings for protons and low energies can be explained in a FEG-model whereas data for helium cannot be understood on a similar basis. Finally, at lowest energies, the agreement with SRIM is found to be worst, which is no surprise due to the lack of experimental data at low energies, this points to the necessity of future experiments at even lower energies, to permit accurate predictions of relevance for ion-materials interaction at commonly employed sputtering energies.

**Acknowledgments**

The authors would like to acknowledge S. Droulias, S. Cederberg and J. Åström for the support with the operation of the accelerators in Sweden. Financial support from Swedish agencies VR-RFI (contracts #821-2012-5144 and #2017-00646_9) and Foundation for Strategic Research (SSF - contract #RIF14-0053) are gratefully acknowledged.

**References**

[1] P. Curie, M. Curie, Sur la charge électrique des rayons déviables du radium, Comptes Rendus - Acad Sci Paris. 130 (1900) 647–650.

[2] S. Creutzburg, E. Schmidt, P. Kutza, R. Loetzsch, I. Uschmann, A. Undisz, M. Rettenmayr, F. Gala, G. Zollo, A. Boulle, A. Debelle, E. Wendler, Defects and mechanical properties in weakly damaged Si ion implanted GaAs, Phys. Rev. B. 99 (2019) 245205. https://doi.org/10.1103/PhysRevB.99.245205.

[3] C. Jeynes, J.L. Colaux, Thin film depth profiling by ion beam analysis, Analyst. 141 (2016) 5944–5985. https://doi.org/10.1039/c6an01167e.

[4] W.J. Weber, Y. Zhang, Predicting damage production in monoatomic and multi-elemental targets using stopping and range of ions in matter code: Challenges and recommendations,

Curr. Opin. Solid State Mater. Sci. 23 (2019) 100757. https://doi.org/10.1016/j.cossms.2019.06.001.

[5] T. Thien Tran, L. Jablonka, C. Lavoie, Z. Zhang, D. Primetzhofer, In-situ characterization of ultrathin nickel silicides using 3D medium-energy ion scattering, (n.d.). https://doi.org/10.1038/s41598-020-66464-1.

[6] O. Jäkel, Physical advantages of particles: protons and light ions, Br. J. Radiol. 93 (2020) 20190428. https://doi.org/10.1259/bjr.20190428.

[7] P. de Vera, R. Garcia-Molina, I. Abril, Simulation of the energy spectra of swift light ion beams after traversing cylindrical targets: a consistent interpretation of experimental data relevant for hadron therapy, Eur. Phys. J. D. 73 (2019). https://doi.org/10.1140/epjd/e2019-100083-4.

[8] R.J. Cortez, J.T. Cassibry, Stopping power in D6Li plasmas for target ignition studies, Nucl. Fusion. 58 (2018) 026009. https://doi.org/10.1088/1741-4326/AA92DE.

[9] B. He, X.J. Meng, Z.G. Wang, J.G. Wang, Ab initio research of energy loss for energetic protons in solid-density Be, Phys. Plasmas. 24 (2017) 033110. https://doi.org/10.1063/1.4977915.

[10] S. Lohmann, D. Primetzhofer, Disparate Energy Scaling of Trajectory-Dependent Electronic Excitations for Slow Protons and He Ions, Phys. Rev. Lett. 124 (2020) 96601. https://doi.org/10.1103/PhysRevLett.124.096601.

[11] F. Matias, R.C. Fadanelli, P.L. Grande, N.R. Arista, N.E. Koval, G. Schiwietz, Stopping power of cluster ions in a free-electron gas from partial-wave analysis, Phys. Rev. A. 98 (2018) 062716. https://doi.org/10.1103/PhysRevA.98.062716.

[12] A.E. Sand, Incorporating Electronic Effects in Molecular Dynamics Simulations of Neutron and Ion-Induced Collision Cascades, in: Handb. Mater. Model., Springer International Publishing, 2020: pp. 2413–2436. https://doi.org/10.1007/978-3-319-44680-6_135.

[13] E.E. Quashie, A.A. Correa, Electronic stopping power of protons and alpha particles in nickel, Phys. Rev. B. 98 (2018) 235122. https://doi.org/10.1103/PhysRevB.98.235122.

[14] E. Gruber, R.A. Wilhelm, R. Pétuya, V. Smejkal, R. Kozubek, A. Hierzenberger, B.C. Bayer, I. Aldazabal, A.K. Kazansky, F. Libisch, A. V. Krasheninnikov, M. Schleberger, S. Facsko, A.G. Borisov, A. Arnau, F. Aumayr, Ultrafast electronic response of graphene to a strong and localized electric field, Nat. Commun. 7 (2016) 1–7. https://doi.org/10.1038/ncomms13948.

[15] E. Fermi, E. Teller, The capture of negative mesotrons in matter, Phys. Rev. 72 (1947) 399–408. https://doi.org/10.1103/PhysRev.72.399.

[16] I. Nagy, A. Arnau, P.M. Echenique, Nonlinear stopping power and energy-loss straggling of an interacting electron gas for slow ions, 1989.

[17] A. Mann, W. Brandt, Material dependence of low-velocity stopping powers, Phys. Rev. B. 24 (1981) 4999–5003. https://doi.org/10.1103/PhysRevB.24.4999.

[18] H. Bethe, J. Ashkin, Experimental nuclear physics. volume I, John Wiley & sons, New York ;London, 1953.

[19] H. Bichsel, Straggling in thin silicon detectors, Rev. Mod. Phys. 60 (1988) 663–699.

https://doi.org/10.1103/RevModPhys.60.663.

[20]  P. Sigmund, Six decades of atomic collisions in solids, Nucl. Instruments Methods Phys. Res. Sect. B Beam Interact. with Mater. Atoms. 406 (2017) 391–412. https://doi.org/10.1016/j.nimb.2016.12.004.

[21]  M. Mayer, S. Möller, M. Rubel, A. Widdowson, S. Charisopoulos, T. Ahlgren, E. Alves, G. Apostolopoulos, N.P. Barradas, S. Donnelly, S. Fazinić, K. Heinola, O. Kakuee, H. Khodja, A. Kimura, A. Lagoyannis, M. Li, S. Markelj, M. Mudrinic, P. Petersson, I. Portnykh, D. Primetzhofer, P. Reichart, D. Ridikas, T. Silva, S.M. Gonzalez de Vicente, Y.Q. Wang, Ion beam analysis of fusion plasma-facing materials and components: facilities and research challenges, Nucl. Fusion. 60 (2019) 025001. https://doi.org/10.1088/1741-4326/ab5817.

[22]  D. Roth, B. Bruckner, M. V. Moro, S. Gruber, D. Goebl, J.I. Juaristi, M. Alducin, R. Steinberger, J. Duchoslav, D. Primetzhofer, P. Bauer, Electronic Stopping of Slow Protons in Transition and Rare Earth Metals: Breakdown of the Free Electron Gas Concept, Phys. Rev. Lett. 118 (2017) 1–5. https://doi.org/10.1103/PhysRevLett.118.103401.

[23]  M.J. Berger, M. Inokuti, H.H. Andersen, H. Bichsel, D. Powers, S.. M. Seltzer, D.. Thwaites, D.E. Watt, Report 49, Stopping Powers and Ranges for Protons and Alpha Particles, Oxford Academic, 1993. https://doi.org/10.1093/JICRU/OS25.2.REPORT49.

[24]  J.F. Ziegler, SRIM - The Stopping and Range of Ions in Matter, (2020). http://www.srim.org/ (accessed July 2, 2020).

[25]  C.C. Montanari, D.M. Mitnik, C.D. Archubi, J.E. Miraglia, Energy loss of protons in W using fully relativistic calculations and mean excitation energies of W, Au, Pb, and Bi, Phys. Rev. A - At. Mol. Opt. Phys. 80 (2009) 1–6. https://doi.org/10.1103/PhysRevA.80.012901.

[26]  P.L. Grande, G. Schiwietz, Impact-parameter dependence of the electronic energy loss of fast ions, Phys. Rev. A - At. Mol. Opt. Phys. 58 (1998) 3796–3801. https://doi.org/10.1103/PhysRevA.58.3796.

[27]  A. Schinner, P. Sigmund, DPASS code - Version 21.06, (2020). https://www.sdu.dk/en/dpass (accessed July 6, 2020).

[28]  D. Roth, C.E. Celedon, D. Goebl, E.A. Sanchez, B. Bruckner, R. Steinberger, J. Guimpel, N.R. Arista, P. Bauer, Systematic analysis of different experimental approaches to measure electronic stopping of very slow hydrogen ions, Nucl. Instruments Methods Phys. Res. Sect. B Beam Interact. with Mater. Atoms. 437 (2018) 1–7. https://doi.org/10.1016/j.nimb.2018.09.028.

[29]  M. V. Moro, R. Holeňák, L. Zendejas Medina, U. Jansson, D. Primetzhofer, Accurate high-resolution depth profiling of magnetron sputtered transition metal alloy films containing light species: A multi-method approach, Thin Solid Films. 686 (2019). https://doi.org/10.1016/j.tsf.2019.137416.

[30]  J.A. Leavitt, L.C. McIntyre, M.D. Ashbaugh, J.G. Oder, Z. Lin, B. Dezfouly-Arjomandy, Cross sections for 170.5° backscattering of 4He from oxygen for 4He energies between 1.8 and 5.0 MeV, Nucl. Inst. Methods Phys. Res. B. 44 (1990) 260–265. https://doi.org/10.1016/0168-583X(90)90637-A.

[31]  A.F. Gurbich, SigmaCalc recent development and present status of the evaluated cross-

sections for IBA, Nucl. Instruments Methods Phys. Res. Sect. B Beam Interact. with Mater. Atoms. 371 (2016) 27–32. https://doi.org/10.1016/j.nimb.2015.09.035.

[32]  P.M. Wolf, B. Bruckner, D. Primetzhofer, Electronic energy loss of light ions in self-supporting foils in transmission and backscattering geometry, Uppsala University, 2019. urn:nbn:se:uu:diva-403474.

[33]  D. Primetzhofer, P. Bauer, Trace element quantification in high-resolution Rutherford backscattering spectrometry, Nucl. Instruments Methods Phys. Res. Sect. B Beam Interact. with Mater. Atoms. 269 (2011) 1284–1287. https://doi.org/10.1016/j.nimb.2010.11.028.

[34]  S.A. Corrêa, E. Pitthan, M. V. Moro, D. Primetzhofer, A multipurpose set-up using keV ions for nuclear reaction analysis, high-resolution backscattering spectrometry, low-energy PIXE and in-situ irradiation experiments, Nucl. Instruments Methods Phys. Res. Sect. B Beam Interact. with Mater. Atoms. 478 (2020) 104–110. https://doi.org/10.1016/j.nimb.2020.05.023.

[35]  M.K. Linnarsson, A. Hallén, J. Åström, D. Primetzhofer, S. Legendre, G. Possnert, New beam line for time-of-flight medium energy ion scattering with large area position sensitive detector, Rev. Sci. Instrum. 83 (2012). https://doi.org/10.1063/1.4750195.

[36]  J. Tesmer, M. Nastasi, Handbook of Modern Ion Beam Materials Analysis, 1st ed, Materials Research Society, Warrendale, 1995.

[37]  W.K. Chu, J.M. Mayer, M.A. Nicolet, Backscattering spectrometry, 1st ed, Academic Press INC, San Diego, 1978.

[38]  S.Z. Izmailov, E.I. Sirotinin, A.F. Tulinov, Energy loss of protons in Si, Ge and Mo, Nucl. Instruments Methods. 168 (1980) 81–84. https://doi.org/10.1016/0029-554X(80)91235-5.

[39]  F.G. Neshev, A.A. Puzanov, K.S. Shyshkin, The determination of energy losses of nitrogen ions from the backscattering spectra, Radiat. Eff. 25 (1975) 271–273. https://doi.org/10.1080/00337577508235400.

[40]  D. Roth, D. Goebl, D. Primetzhofer, P. Bauer, A procedure to determine electronic energy loss from relative measurements with TOF-LEIS, Nucl. Instruments Methods Phys. Res. Sect. B Beam Interact. with Mater. Atoms. 317 (2013) 61–65. https://doi.org/10.1016/j.nimb.2012.12.094.

[41]  J.P. Biersack, E. Steinbauer, P. Bauer, A particularly fast TRIM version for ion backscattering and high energy ion implantation, Nucl. Inst. Methods Phys. Res. B. 61 (1991) 77–82. https://doi.org/10.1016/0168-583X(91)95564-T.

[42]  M. Mayer, SIMNRA, a simulation program for the analysis of NRA, RBS and ERDA, in: AIP Conf. Proc., AIP Publishing, 2008: pp. 541–544. https://doi.org/10.1063/1.59188.

[43]  S.N. Markin, D. Primetzhofer, M. Spitz, P. Bauer, Electronic stopping of low-energy H and He in Cu and Au investigated by time-of-flight low-energy ion scattering, Phys. Rev. B - Condens. Matter Mater. Phys. 80 (2009) 205105. https://doi.org/10.1103/PhysRevB.80.205105.

[44]  H. Paul, D. Semrad, A. Seilinger, Reference stopping cross sections for hydrogen and helium ions in selected elements, Nucl. Inst. Methods Phys. Res. B. 61 (1991) 261–281. https://doi.org/10.1016/0168-583X(91)95630-V.

[45]  H. Paul, A. Schinner, Judging the reliability of stopping power tables and programs for protons and alpha particles using statistical methods, Nucl. Instruments Methods Phys. Res. Sect. B Beam Interact. with Mater. Atoms. 227 (2005) 461–470. https://doi.org/10.1016/j.nimb.2004.10.007.

[46]  H. Paul, D. Sánchez-Parcerisa, A critical overview of recent stopping power programs for positive ions in solid elements, Nucl. Instruments Methods Phys. Res. Sect. B Beam Interact. with Mater. Atoms. 312 (2013) 110–117. https://doi.org/10.1016/j.nimb.2013.07.012.

[47]  H. Paul, Recent results in stopping power for positive ions, and some critical comments, Nucl. Instruments Methods Phys. Res. Sect. B Beam Interact. with Mater. Atoms. 268 (2010) 3421–3425. https://doi.org/10.1016/j.nimb.2010.09.001.

[48]  Electronic Stopping Power of Matter for Ions, (2020). https://www-nds.iaea.org/stopping/ (accessed July 1, 2020).

[49]  M. V. Moro, B. Bruckner, P.L. Grande, M.H. Tabacniks, P. Bauer, D. Primetzhofer, Stopping cross section of vanadium for H+ and He+ ions in a large energy interval deduced from backscattering spectra, Nucl. Instruments Methods Phys. Res. Sect. B Beam Interact. with Mater. Atoms. 424 (2018) 43–51. https://doi.org/10.1016/j.nimb.2018.03.032.

[50]  M.A. Sortica, M.K. Linnarsson, D. Wessman, S. Lohmann, D. Primetzhofer, A versatile time-of-flight medium-energy ion scattering setup using multiple delay-line detectors, Nucl. Instruments Methods Phys. Res. Sect. B Beam Interact. with Mater. Atoms. 463 (2020) 16–20. https://doi.org/10.1016/j.nimb.2019.11.019.

[51]  E.A. Figueroa, N.R. Arista, J.C. Eckardt, G.H. Lantschner, Determination of the difference between the mean and the most probable energy loss of low-energy proton beams traversing thin solid foils, Nucl. Instruments Methods Phys. Res. Sect. B Beam Interact. with Mater. Atoms. 256 (2007) 126–130. https://doi.org/10.1016/j.nimb.2006.11.103.

[52]  B. Bruckner, D. Roth, D. Goebl, P. Bauer, D. Primetzhofer, A note on extracting electronic stopping from energy spectra of backscattered slow ions applying Bragg's rule, Nucl. Instruments Methods Phys. Res. Sect. B Beam Interact. with Mater. Atoms. 423 (2018) 82–86. https://doi.org/10.1016/j.nimb.2018.02.005.

[53]  W.H. Bragg, R. Kleeman,  XXXIX. On the α particles of radium, and their loss of range in passing through various atoms and molecules , London, Edinburgh, Dublin Philos. Mag. J. Sci. 10 (1905) 318–340. https://doi.org/10.1080/14786440509463378.

[54]  JCGM, Evaluation of measurement data-Guide to the expression of uncertainty in measurement Évaluation des données de mesure-Guide pour l'expression de l'incertitude de mesure, 2008. www.bipm.org.

[55]  JCGM 102:2011, Evaluation of measurement data – Supplement 2 to the "Guide to the expression of uncertainty in measurement" – Extension to any number of output quantities, Jcgm. 102 (2011) 1–72.

[56]  J.F. Ziegler, J.P. Biersack, U. Littmark, The stopping and range of ions in solids, 2nd ed, New York, NY, 1985.

[57]  M. Luomajarvi, Stopping powers of some metals for 0.3-1.5 MeV protons, Radiat. Eff. 40 (1979) 173–179. https://doi.org/10.1080/00337577908237920.

[58] E.I. Sirotinin, A.F. Tulinov, V.A. Khodyrev, V.N. Mizgulin, Proton energy loss in solids, Nucl. Inst. Methods Phys. Res. B. 4 (1984) 337–345. https://doi.org/10.1016/0168-583X(84)90577-9.

[59] V.Y. Chumanov, S.Z. Izmailov, G.P. Pokhil, E.I. Sirotinin, A.F. Tulinov, On the determination of energy losses by charged particles from the backscattered energy spectra, Phys. Status Solidi. 53 (1979) 51–62. https://doi.org/10.1002/pssa.2210530104.

[60] E.I. Strotinin, A.F. Tulinov, A. Fiderkevich, K.S. Shyshkin, The determination of energy losses from the spectrum of particles scattered by a thick target, Radiat. Eff. 15 (1972) 149–152. https://doi.org/10.1080/00337577208234688.

[61] E. Leminen, A. Fontell, STOPPING POWER OF Ti, Mo, Ag, Ta AND W FOR 0. 5-1. 75 MeV 4He IONS., Radiat. Eff. 22 (1974) 39–44. https://doi.org/10.1080/00337577408232143.

[62] J.A. Borders, HELIUM ION STOPPING CROSS SECTIONS IN BISMUTH, LEAD AND TUNGSTEN., Radiat. Eff. 21 (1974) 165–169. https://doi.org/10.1080/00337577408241458.

[63] W.K. Chu, J.F. Ziegler, I. V. Mitchell, W.D. MacKintosh, Energy-loss measurements of 4He ions in heavy metals, Appl. Phys. Lett. 22 (1973) 437–439. https://doi.org/10.1063/1.1654703.

[64] W.K. Lin, H.G. Olson, D. Powers, Alpha-particle stopping cross section of solids from 0.3 to 2.0 MeV, Phys. Rev. B. 8 (1973) 1881–1888. https://doi.org/10.1103/PhysRevB.8.1881.

[65] N. National Institute of Standards and Technology, The PSTAR program, (2020). https://physics.nist.gov/PhysRefData/Star/Text/PSTAR.html (accessed July 2, 2020).

[66] N. National Institute of Standards and Technology, The ASTAR program, (2020). https://physics.nist.gov/PhysRefData/Star/Text/ASTAR.html (accessed July 2, 2020).

[67] U. Fano, Inelastic collisions and the molière theory of multiple scattering, Phys. Rev. 93 (1954) 117–120. https://doi.org/10.1103/PhysRev.93.117.

[68] M. Inokuti, Inelastic collisions of fast charged particles with atoms and molecules-The bethe theory revisited, Rev. Mod. Phys. 43 (1971) 297–347. https://doi.org/10.1103/RevModPhys.43.297.

[69] J. Fuhr, V. Ponce, F. García de Abajo, Dynamic screening of fast ions moving in solids, Phys. Rev. B - Condens. Matter Mater. Phys. 57 (1998) 9329–9335. https://doi.org/10.1103/PhysRevB.57.9329.

[70] G. Schiwietz, P.L. Grande, Unitary convolution approximation for the impact-parameter dependent electronic energy loss, Nucl. Instruments Methods Phys. Res. Sect. B Beam Interact. with Mater. Atoms. 153 (1999) 1–9. https://doi.org/10.1016/S0168-583X(98)00981-1.

[71] L. De Ferrariis, N.R. Arista, Classical and quantum-mechanical treatments of the energy loss of charged particles in dilute plasmas, Phys. Rev. A. 29 (1984) 2145–2159. https://doi.org/10.1103/PhysRevA.29.2145.

[72] F. Matias, P.L. Grande, M. Vos, P. Koval, N.E. Koval, N.R. Arista, Nonlinear stopping effects of slow ions in a no-free-electron system: Titanium nitride, Phys. Rev. A. 100 (2019) 030701. https://doi.org/10.1103/PhysRevA.100.030701.

[73] P. Sigmund, A. Schinner, Binary stopping theory for swift heavy ions, Eur. Phys. J. D. 8 (2000) 425–434. https://doi.org/10.1140/epjd/e2005-00323-2.

[74] N. Bohr, On the theory of the decrease of velocity of moving electrified particles on passing through matter: Phil. mag. 25 (1913) 10–31, Niels Bohr Collect. Work. 8 (1987) 47–71. https://doi.org/10.1016/S1876-0503(08)70140-3.

[75] A. Schinner, P. Sigmund, Expanded PASS stopping code, Nucl. Instruments Methods Phys. Res. Sect. B Beam Interact. with Mater. Atoms. 460 (2019) 19–26. https://doi.org/10.1016/j.nimb.2018.10.047.

[76] C.C. Montanari, C.D. Archubi, D.M. Mitnik, J.E. Miraglia, Energy loss of protons in Au, Pb, and Bi using relativistic wave functions, Phys. Rev. A - At. Mol. Opt. Phys. 79 (2009) 1–9. https://doi.org/10.1103/PhysRevA.79.032903.

[77] A.E. Sand, K. Nordlund, On the lower energy limit of electronic stopping in simulated collision cascades in Ni, Pd and Pt, J. Nucl. Mater. 456 (2015) 99–105. https://doi.org/10.1016/j.jnucmat.2014.09.029.

[78] J. Tian, W. Zhou, Q. Feng, J. Zheng, Molecular dynamics simulations with electronic stopping can reproduce experimental sputtering yields of metals impacted by large cluster ions, Appl. Surf. Sci. 435 (2018) 65–71. https://doi.org/10.1016/j.apsusc.2017.11.080.

[79] P.M. Echenique, R.M. Nieminen, R.H. Ritchie, Density functional calculation of stopping power of an electron gas for slow ions, Solid State Commun. 37 (1981) 779–781. https://doi.org/10.1016/0038-1098(81)91173-X.

[80] J.I. Juaristi, M. Alducin, R. Díez Muiño, H.F. Busnengo, A. Salin, Role of Electron-Hole Pair Excitations in the Dissociative Adsorption of Diatomic Molecules on Metal Surfaces, (2007). https://doi.org/10.1103/PhysRevLett.100.116102.

[81] D. Isaacson, Compilation of rs Values (Internal Report - Radiation and Solid State Laboratory), New York University, 1975.

[82] M. V. Moro, P. Bauer, D. Primetzhofer, Experimental electronic stopping cross section of transition metals for light ions – systematics around the stopping maximum, Phys. Rev. A - At. Mol. Opt. Phys. (Accepted (2020).

[83] D. Primetzhofer, S. Rund, D. Roth, D. Goebl, P. Bauer, Electronic excitations of slow ions in a free electron gas metal: Evidence for charge exchange effects, Phys. Rev. Lett. 107 (2011) 1–5. https://doi.org/10.1103/PhysRevLett.107.163201.

[84] S. Rund, D. Primetzhofer, S.N. Markin, D. Goebl, P. Bauer, Charge exchange of He+-ions with aluminium surfaces, Nucl. Instruments Methods Phys. Res. Sect. B Beam Interact. with Mater. Atoms. 269 (2011) 1171–1174. https://doi.org/10.1016/j.nimb.2010.11.049.

[85] F. Matias, R.C. Fadanelli, P.L. Grande, N.E. Koval, R.D. Muiño, A.G. Borisov, N.R. Arista, G. Schiwietz, Ground- and excited-state scattering potentials for the stopping of protons in an electron gas, J. Phys. B At. Mol. Opt. Phys. 50 (2017). https://doi.org/10.1088/1361-6455/aa843d.

[86] R.A. Wilhelm, P.L. Grande, Unraveling energy loss processes of low energy heavy ions in 2D materials, Commun. Phys. 2 (2019) 1–8. https://doi.org/10.1038/s42005-019-0188-7.